\documentstyle[12pt]{article}
\date{November 7th 2006}
\begin{document}
\title{\begin{flushright}{\small IFUM-880-FT}\end{flushright}
Gauge boson families in grand unified theories of fermion 
masses: 
E$_6^4 >{\hspace{-8pt}}\triangleleft$ S$_4$}
\author{{\small Francesco Caravaglios and Stefano Morisi} \\
{\small Dipartimento di Fisica, Universit\`{a} di Milano, Via Celoria 16,
I-20133 Milano, Italy} \\
{\small and}\\
{\small INFN,\ sezione di Milano}}
\maketitle

\begin{abstract}
In third quantization the origin of fermion families is easy to understand:
the electron field, the muon field and the tau field 
are identical fields in precisely the  same  sense as three electrons 
are identical and indistinguishable particles of a theory of second 
quantization.
In both cases, the permutation of these fields or particles
  leaves the lagrangian invariant. 
One can also extend the concept of family to gauge bosons. This can be
obtained through the semidirect product of the  gauge group with the
group of permutations of $n$ objects. In this paper we have studied 
the group    E$_6^4 >{\hspace{-6pt}}\triangleleft$ S$_4$.  We  
explain why we have chosen $E_6$ as fundamental gauge group factor and 
why we start with a model with four gauge boson/fermion families  to 
accommodate and to 
 fit  the  standard model with only three fermion families.
We suggest a possible symmetry breaking pattern of  
E$_6^4 >{\hspace{-6pt}}\triangleleft$ S$_4$ that could explain 
 quark, lepton and neutrino masses and mixings. 
\end{abstract}
\subsection*{Introduction}
Recent neutrino experiments \cite{exp1} have shown oscillations among
 neutrinos of 
different flavor.
This is due to the fact that these neutrinos have masses and mixing angles 
between different generations.
Differently from the quark sector, where all mixing angles are small,
in neutrino oscillations some angles are large or even maximal, like in 
atmospheric neutrinos.
Since in grand unified theories there are stringent relations due to 
the grand unified symmetry between the quark sector and the lepton sector, it 
is necessary to understand quark/lepton  differences
 within grand unified models.
For instance, in the simplest SU(5) models we find that the down quark 
mass matrix is the transpose of the charged lepton mass matrix. In SO(10) 
 the up quark Yukawa  matrix is equal to the Dirac   matrix of the neutrinos.

Several recent works have found interesting explanations of  
the amazing experimental results 
on neutrino oscillations,  extending the standard model with new discrete 
 flavor symmetries in the lepton sector 
\cite{s3,Morisi:2006pf,Caravaglios:2005gw,Caravaglios:2005gf,
Fritzsch:2004xc,A4,mutau,s4}.
The advantage of the discrete symmetries with respect to the continuous one,
is that they can naturally explain the peculiar structure of large mixings 
 with a relatively large hierarchy between atmospheric and solar neutrino 
masses.  
For instance, a neutrino mass matrix that is almost exactly invariant 
under the permutations of the two heaviest families, that is a permutation 
of $\tau$ neutrino with the $\mu$ neutrino, directly predicts a maximal 
 atmospheric  angle and $\theta_{13}\simeq 0$. In perfect agreement 
with the experimental data. In fact if P is the operator that permutates
the two heaviest families and P commutes with the mass matrix, 
then it exists  an  eigenvector of P $(0,-1/\sqrt{2},1/\sqrt{2})$,
 with eigenvalue -1,  and this is also a mass eigenstate. 
If, for instance, this 
is the heaviest neutrino state then   $\theta_{13}= 0$ and the atmospheric 
 angle is maximal.
Here we have assumed a diagonal charged lepton mass matrix. This is not always 
true, and it seems to be in contradiction with the fact that left-handed 
leptons belong to the same SU(2) doublet (see the section on the
 $S_3$ breaking misalignment).  
In a recent paper \cite{Caravaglios:2005gw}
 we have shown that including such a $S_2$ symmetry in a 
larger discrete symmetry $S_3$ that permutates all three families, also 
the solar mixing angle can be explained in a natural way.  It is important 
to note that such a discrete symmetry directly comes from an extension 
of field theory, that gives an explanation of the  hierarchy between 
the unification scale and the electroweak scale , this is the 
 so-called third  quantization \cite{Caravaglios:2002ws}.
 
The difficulty that we have mentioned above is how to embed 
such a $S_3$ symmetry in a grand unified model, being compatible 
with the experimental observation of a strong hierarchy between physical 
quark masses 
and small mixing angles.

Several authors are studying this problem, but  some questions 
 are still open, and in some cases a fine-tuning is required to 
make leptons and quark results  compatible. 

In this paper we make a step forward in the grand unified model building 
with respect a previous work \cite{Caravaglios:2005gf}. 
In fact we give a more convincing explanation 
of the reason why the $S_3$ singlet is the lightest among 
 the  left-handed neutrinos and also why there is  so large splitting 
between the two components of the $S_3$ doublet.
 In addition we ameliorate the study of 
the $E_6$ unified group \cite{Slansky:1981yr,E6} embedding it in 
E$_6^4 >{\hspace{-6pt}}\triangleleft$ S$_4$ , and giving a symmetry breaking 
pattern that simultaneously explains quark and lepton masses and mixings.

At the beginning we will discuss which kind of mass operators that break the 
$S_3$ symmetry are  necessary  in order to explain the quark 
 mass and mixings. To do this we will make use of a montecarlo statistical 
analysis of the allowed mass operators, the method and  some 
 results can be found  in \cite{Morisi:2006pf}.
Secondly we will extend our analysis to the charged lepton sector, putting 
in evidence the difficulties due to the grand unification hypothesis and how
to overcome them.  Next we propose a model of unification
 E$_6^4 >{\hspace{-6pt}}\triangleleft$ S$_4$ 
that is the semidirect product of the gauge group 
  $E_6^4$ with the  flavor group $S_4$ \cite{s4}, and one possible breaking 
pattern.
Note that the gauge group $E_6^4$  does not commute with $S_4$, but it is 
the normal subgroup of the group $G$ obtained as the semidirect product of
 $E_6^4$ with $S_4$, the permutation symmetry of four families. In such a 
model  the concept of family is extended also to gauge bosons.
Each family is composed of a subset of fermions and their own gauge bosons 
that mediate interactions between the fermion of one family only. Namely,
in our specific model, the first family contains 27 fermions 
(the smallest $E_6$ irreducible representation) and 78 gauge bosons 
(the vector boson of $E_6$). These bosons only interact   with the 27 
fermions  of the first family, while fermions of the other families are 
neutral (with respect them). Similarly the 78 gauge bosons of the second
 family (corresponding to the second $E_6$ factor  in $E_6^4$) only 
interacts with the fermions of the second family. In other words 
the  $S_4$ symmetry group permutates the four families, thus it permutates 
 not only the fermions of the four families but also their gauge bosons.
 This is another 
way to understand why $S_4$ does not commute with $E_6^4$,
and why $G$ is the semidirect product ( and not direct) between  $E_6^4$ and 
 $S_4$. 
At the end we will see how this $G$ group can be broken in order 
to explain in a natural way  masses and mixings both 
of quarks and leptons.
\subsection*{The  flavor symmetry and   gauge unification}
Putting together the flavor symmetry and gauge unification is not trivial.
After the experimental observation of large mixing angles in neutrino 
oscillations  a  distinction between quarks, and leptons 
has become manifest.
Small mixing angles and large mass hierarchies in the quark sector do not 
show any apparent symmetry.
On the contrary neutrino oscillations are due to  a peculiar mixing matrix 
with two large mixing angles, $\sin^2(\theta_{12})$ and $\sin^2(\theta_{23})$,
from the solar and atmospheric neutrinos. 
   $\sin^2(\theta_{13})$ is very 
small and close to zero.
This pattern clearly demands  an explanation. As already mentioned in the 
introduction,  an almost $S_3$  symmetric right-handed  
 neutrino mass matrix is diagonalized by a unitary matrix $V_\nu$ 
that is very similar 
to that one required to explain all neutrino experiments.      
However there are some issues that come out 
 when one want to apply this symmetry 
to the full lagrangian and in particular if one also requires grand 
unification. 

First, if we want the matrix  $V_\nu$  to be the one describing 
neutrino oscillations, {\it i.e.}  $V_{\mathrm{PMNS}}=V_\nu$, we need
both an $S_3$ symmetric Dirac neutrino mass matrix and  a 
diagonal mass matrix for the charged leptons.
 Due to the large lepton  mass 
hierarchies this requirement is incompatible with a $S_3$ symmetric
charged lepton  mass matrix.    
At the same time, the left-handed component of charged leptons belong to the
  same $SU(2)$ electroweak doublet of  left-handed neutrinos. 
Both fermions must satisfy the same transformation properties under the 
$S_3$ symmetry. 
We infer that a diagonal charged lepton yukawa interaction can appear only 
after the $S_3$ symmetry is fully broken.  
A similar argument lead to the same conclusions  for quarks.
In  grand unified models this means that we have to go beyond the minimal 
scenario. 

For example,   in the simplest   $E_6$  grand unification both 
the fermion families $27_i^\alpha$ and the scalar electroweak doublet 
 $\tilde {27}^\gamma$ live in the same ({\bf 27})
 irreducible representation. Only one renormalizable  yukawa interaction
is allowed by symmetry arguments 
\begin{equation}
g_{ij}\, 27_i^\alpha \, 27_j^\beta \, \tilde {27}^\gamma \, 
\varepsilon_{\alpha\beta\gamma}.   
\end{equation}
This yukawa gives the relation
\begin{equation}
M_e=M_d=M_\nu=M_u \label{mat}
\end{equation}
where $M_\nu$ is the dirac neutrino mass matrix. The relations (\ref{mat}) 
imply that if $M_\nu$ satisfies a symmetry requirement the same must be true 
for the other  fermion matrices.
To avoid these problematic relations we go beyond the minimal $E_6$ model.
Namely if we choose the Higgs doublet $H$ 
 in the $\overline{16}\in 351^\prime$ of 
 $SO(10)\subset E_6$, the yukawa interaction becomes 
\begin{equation}
g_{ij}\, 27_i^\alpha \, 27_j^\beta \, \tilde {351}^\prime_{\alpha\beta}   
\end{equation}
and we get\footnote{In this model, the right-handed neutrino is the $SO(10)$ 
singlet contained in the {\bf 27 } of $E_6$. See \cite{Caravaglios:2005gf}.} 
\begin{equation}
M_\nu^{ij}=g_{ij} H 
\end{equation}
\begin{equation}
M_e=M_d=M_u=0.
\end{equation}
Now only $M_\nu$ appears at the tree level  in the fundamental lagrangian and 
it must 
 satisfy all symmetry requirements. $ M_e,M_d,M_u$ will appear after the 
$S_3$ is fully broken and are induced by radiative corrections\footnote{All 
standard model  yukawa couplings are very small, and this is compatible 
with this scenario. Only the top quark has a large yukawa coupling, the reason 
could be due to a non perturbative correction, like those  
coming from  the renormalization  group evolution.}.
Now we have seen that  the choice of a particular representation for the Higgs 
doublet give different mass matrices, and for instance $M_\nu\neq M_e$.

This argument can explain why  $M_\nu$ is $S_3$ symmetric, while $M_e$ is not 
$S_3$ symmetric, but if we want to go further to explain atmospheric neutrino 
oscillations and $\sin^2\theta_{13}\simeq 0$ we  have to break the $S_3$ 
symmetry  in the neutrino sector (as already mentioned in the introduction).
 This breaking occurs in a different 
direction with respect to the charged lepton sector. 
Neutrino data suggest that $S_3$ is broken into a $S_2$, that exchanges the
 $\tau $ neutrino with the $\mu$ neutrino. On the contrary 
 the $\tau$ lepton is the heaviest 
and this  suggests a breaking of $S_3$ that leaves the exchange symmetry
 between the muon and the electron. Thus we have also to understand why the 
direction of the breaking of $S_3$ shows a difference between 
  neutrinos and charged leptons. This issue will become more clear after when 
 we will  discuss  with an explict example the origin of this difference.

\subsection*{Quark masses}
Before  discussing grand unified models we want  to understand 
how to break $S_3$ in the quark sector, namely which operators we need 
to introduce in order to explain quark masses and mixings.
If we want to extend the standard model in a theory of third quantization, 
that  explains why we have replications of   fermion families, then we have
 to assume that the three fermion families transform as a triplet 
with  respect $S_3$  \cite{Caravaglios:2002ws}. 
This triplet  is a reducible representation  of $S_3$,
and more precisely it is the sum of a doublet and a singlet of $S_3$. 
This is a natural choice in the context of third quantization:
second quantization is a theory that imposes that all electrons 
are identical particles, that is the hamiltonian that describes the
time evolution of the physical states of $n$ electrons 
 must be symmetric under permutations 
of the space coordinates of the $n$ electron positions.
The analogue of this picture in third quantization is a system of $n$
 fermion families that is described by an hamiltonian that is symmetric under 
permutations $S_n$ among  these fermion families.  As a consequence in a 
world with  three  families , these families must transform as a triplet of 
$S_3$. In our world  this symmetry is not exact, otherwise we would observe 
two charged leptons degenerate in mass. To spontaneously break this 
symmetry $S_3$ we will introduce a scalar field that is a standard model 
singlet and a triplet  with respect $S_3$.  Also in this case the choice of 
a $S_3$ triplet instead of a doublet comes from third quantization.  
We will assume that the three components of this field $\phi_i$ (with $i=1,3$)
take three different vev $v_3 \gg v_2\gg v_1$ in order to completely break 
the group $S_3$, while  the Higgs responsible for the electroweak breaking 
is a singlet of $S_3$. In fact we will see that even if we choose 
the electroweak Higgs to be a triplet under $S_3$, at the electroweak scale
 we will  have  only the singlet component.   
It is better to require only one Higgs doublet at the electroweak scale 
to avoid  potentially dangerous flavor changing neutral currents.

Let us start observing that operators like 
\begin{equation}
 \sum_{i,j} \left( g \bar d_L^i d_R^i H+ g^\prime \bar d_L^i d_R^j H
\right) \label{down}
\end{equation}
 that are $S_3$ invariant are disfavored , 
because they give two degenerate masses and large mixing angles. In principle 
one could choose  $g^\prime\gg g$ to obtain one large mass and two light 
states nearly degenerate, but such a hierarchy 
is difficult to understand in the context of a grand unified model with $S_3$ 
as a flavor group. However this possibility is not ruled out 
 and probably deserves further investigation. 
We prefer to get rid of these operators, and for this we will make use 
of an additional $U(1)$ symmetry from the $E_6$ group.
\begin{center}
\begin{table}
$
\begin{array}{ll}
\begin{array}{l}
\hline
{\small \mbox{Fields}~~U_{\mathrm{down}}~~~~Y~~~~I_{3}^{w}~
~~Q_{r}~~~~Q_{t}~~ }\\ \hline \\ 
{\small
\begin{tabular}{cccccc}
${\bar{D}_{1}}$ & $+8$ & $-{\frac{1}{3}}$ & $0$ & $2$ & $-2$  \\ 
$\bar{D}_{2}$ & $+8$ & $-{\frac{1}{3}}$ & $0$ & $2$ & $-2$ \\ 
${\ u_{R3}^{c}}$ & $-4$ & $-{\frac{2}{3}}$ & $0$ & $-1$ & $1$ \\ 
$\bar{D}_{3}$ & $+8$ & $-{\frac{1}{3}}$ & $0$ & $2$ & $-2$  \\ 
$u_{R2}^{c}$ & $-4$ & $-{\frac{2}{3}}$ & $0$ & $-1$ & $1$  \\ 
$u_{R1}^{c}$ & $-4$ & $-{\frac{2}{3}}$ & $0$ & $-1$ & $1$  \\ 
$N_{L}$ & $+8$ & ${\frac{1}{2}}$ & $-{\frac{1}{2}}$ & $2$ & $-2$ \\ 
$d_{L1}$ & $-4$ & ${\frac{1}{6}}$ & $-{\frac{1}{2}}$ & $-1$ & $1$\\ 
$d_{L2}$ & $-4$ & ${\frac{1}{6}}$ & $-{\frac{1}{2}}$ & $-1$ & $1$\\ 
$d_{L3}$ & $-4$ & ${\frac{1}{6}}$ & $-{\frac{1}{2}}$ & $-1$ & $1$\\ 
${El}$ & $+8$ & ${\frac{1}{2}}$ & ${\frac{1}{2}}$ & $2$ & $-2$ \\ 
$u_{L1}$ & $-4$ & ${\frac{1}{6}}$ & ${\frac{1}{2}}$ & $-1$ & $1$ \\ 
${u_{L2}}$ & $-4$ & ${\frac{1}{6}}$ & ${\frac{1}{2}}$ & $-1$ & $1$\\ 
${u_{L3}}$ & $-4$ & ${\frac{1}{6}}$ & ${\frac{1}{2}}$ & $-1$ & $1$\\ 
\end{tabular}
}
\end{array}&
\begin{array}{l}
\hline
{\small \mbox{Fields}~~U_{\mathrm{down}}~~Y~~~~~I_{3}^{w}~~~Q_{r}~~~~Q_{t}~~ }\\ \hline \\ 
{\small
\begin{tabular}{cccccc}
$e_{R}^{c}$ & $-4$ & $1$ & $0$ & $-1$ & $1$ \\ 
${\nu _{R}^{c}}$ & $-8$ & $0$ & $0$ & $-5$ & $1$ \\ 
${D_{1}}$ & $+4$ & ${\frac{1}{3}}$ & $0$ & $-2$ & $-2$  \\ 
${D_{2}}$ & $+4$ & ${\frac{1}{3}}$ & $0$ & $-2$ & $-2$   \\ 
${D_{3}}$ & $+4$ & ${\frac{1}{3}}$ & $0$ & $-2$ & $-2$ \\ 
${\bar{N}_{L}}$ & $+4$ & $-{\frac{1}{2}}$ & ${\frac{1}{2}}$ & $-2$ &$-2$ \\ 
${\bar{E}_{L}}$ & $+4$ & $-{\frac{1}{2}}$ & $-{\frac{1}{2}}$ & $-2$ & $-2$ \\ 
${X_{L}}$ & $-12$ & $0$ & $0$ & $0$ & $4$ \\ 
${d_{R1}^{c}}$ & $0$ & ${\frac{1}{3}}$ & $0$ & $3$ & $1$  \\ 
${d_{R2}^{c}}$ & $0$ & ${\frac{1}{3}}$ & $0$ & $3$ & $1$ \\ 
${d_{R3}^{c}}$ & $0$ & ${\frac{1}{3}}$ & $0$ & $3$ & $1$  \\ 
${\nu _{L}}$ & $0$ & $-{\frac{1}{2}}$ & ${\frac{1}{2}}$ & $3$ & $1$  \\ 
${e_{L}}$ & $0$ & $-{\frac{1}{2}}$ & $-{\frac{1}{2}}$ & $3$ & $1$ \\
&&&&&
\end{tabular}
}
\end{array}\\
&\\
\hline
\end{array}
$
\caption{Quantum numbers of the irrep {\bf 27}} \label{tab1}
\end{table}
\end{center} 
The $E_6$ group contains two extra U(1) that commute with the standard
 model 
gauge group. They can be defined  from the following embedding
E6$\supset$SO(10)$\times$U(1)$_t 
\supset$SU(5)$\times$U(1)$_r\times$U(1)$_t$ (see Table  \ref{tab1}).
If we choose the electroweak Higgs doublet $H$ with 
charges\footnote{this choice is possible if the Higgs belongs to the 
$351^\prime$ 
representation of $E_6$ or the {\bf 1404$_s$} of
E$_6^4 >{\hspace{-6pt}}\triangleleft$ S$_4$,
 see the appendix and \cite{Caravaglios:2005gf,Slansky:1981yr}}
 $(q_r,q_t)=(-3,-5)$,
 the Yukawa couplings in eq.(\ref{down}) are forbidden.
 In this case the 
Yukawa couplings for the quark sector arise as higher mass  dimension 
operators, and only  
after both  the 
 $S_3$ symmetry and  the additional U(1) are broken 
(this explains why this symmetry 
is not observed in the quark masses). To be more specific, 
we can choose the triplet of $S_3$, $\phi_i^d$, responsible for the $S_3$ 
  breaking to have the 
following charges $(q_r,q_t)=(+5,+7)$. Such a field can be found in the 
{\bf 1728} of $E_6$. With these choices, the following 
operators are compatible  both with $S_3$ and U(1)$_r\times $ U(1)$_t$ 
\begin{equation}
\sum_{i,j,k} \left(g^d \bar d_L^i d_R^i H \phi^d_i + g^d_L \bar d_L^i d_R^j H
 \phi^d_i+
g^d_R \bar d_L^i d_R^j H \phi^d_j+g^d_3 \bar d_L^i d_R^j H \phi^d_k\right).
\label{eq7}
\end{equation}
When the fields $\phi_i^d$ will take a vev, the Yukawa 
coupling between the electroweak Higgs and the quarks will appear.
To go forward  with  the model building, we will use a statistical analysis 
where some preferred values for the $g^d$ couplings have been selected. An
 extensive  discussion of the method can be found in \cite{Morisi:2006pf}.
 In particular it has been found that 
$g_L^d$ is slightly smaller than $g^d$, while $g_R^d$ and $g_3^d$ are of the
 same 
order of magnitude and very  small. Furthermore it has been found 
$v_3 \gg v_2 \gg v_1$. The up quark mass matrix is found to be approximately
diagonal.
For this reason we  neglect in the up quark sector  terms proportional to
$g^u_L,g^u_R,g^u_3$, and only one operator will contribute to the up quark 
Yukawa interactions   
\begin{equation}\label{up}
\sum_{i} \, g^u \bar u_L^i u_R^i H \phi^u_i .
\end{equation}

\begin{table}[h]
$$
\begin{tabular}{ll}
\hline
up&down\\
\hline
&\\
$\phi^u_1/\phi^u_3=m_u/m_t$&$\phi^d_1/\phi^d_3=0.000077$\\
$\phi^u_2/\phi^u_3=m_c/m_t$&$\phi^d_2/\phi^d_3=0.054783$\\
$g=1$ & $g=0.848$ \\
$g_L\simeq 0$ & $g_L=0.305~e^{-i~0.02}$\\
$g_R \simeq 0$ & $g_R=0.002~e^{-i~0.12}$\\
$g_3 \simeq 0$ & $g_3~=0.009~e^{i~1.82}$\\
\hline
\end{tabular}
$$
\caption{$g$ couplings in the quark sector. We define $\phi_3=1$ in order 
to set  the mass scale of all $g$ couplings. These have the dimension of 
the inverse of a mass. 
}\label{tab2}
\end{table}


The fact that $g_3^d$ and $g_R^d$ are very small and of the same order of 
magnitude can help us in the construction of the model, since this can 
indirectly   indicate the existence of an unbroken symmetry that forbids 
these operators.
In fact we can extend the standard model with few additional U(1) as follows:
let us choose the following group
$ G=(U(1)_r\times U(1)_t)^3 >{\hspace{-6pt}}\triangleleft S_3$. 
Each $(U(1)_r\times U(1)_t)$ gauge symmetry factor needs two gauge 
bosons that belongs to the $i$-th family.
In practice we enlarge the concept of family to the gauge bosons. 
Each family contains 27 fermions (the irreducible representation 
of $E_6$ ) plus two gauge bosons of the  $(U(1)_r\times U(1)_t)$ group
 relative to the same  family. These two gauge bosons of the $i$-th family 
interact  only with the $i$-th fermion  family\footnote{Next we will see
that each family will contain all the 78 bosons of $E_6$, but 
here we  focus on just these two abelian  factors
 $(U(1)_r\times U(1)_t)$, since the rest of gauge bosons are irrelevant for
the discussion below.}.  These two gauge bosons transform as a triplet with 
respect the permutation group $S_3$. For this reason the group 
$ G=(U(1)_r\times U(1)_t)^3 >{\hspace{-6pt}}\triangleleft S_3$ is 
the semidirect product and not the direct product.
It is easy to see that only the operators below are compatible with G
\begin{equation}
g \bar d_L^i d_R^i H^i \phi_i. 
\end{equation}
Note that we have added the family index $i$ also to the electroweak 
doublet  $H^i$. This is a necessary choice if we want to make a $G$ invariant 
lagrangian.
On the contrary, the operator 
\begin{equation}\label{lr1}
\sum_{i,j} g_L \bar d_L^i d_R^j H^i \phi_i 
\end{equation}
is not G invariant.
In fact for  $j\ne i$, the  $d_R^j$ is the only fermion field that carries 
non zero charges with respect the U(1) of  $j$-th family: the operator 
(\ref{lr1}) does not conserve the charge of the $j$-th family gauge
group.
This justifies  $g \gg g_L$. This  hierarchy is not  large, as  we 
can see in Table \ref{tab2}.
A hierarchy much more important is between $g_L$ and  $g_3,g_R$. To explain 
such a hierarchy we can properly choose the breaking direction of G. 
The only choice consistent with  $g_L \gg g_R,g_3$ is the breaking of 
G in  $G^\prime=(U(1)_{\mathrm{down}})^3 >{\hspace{-6pt}}\triangleleft
 S_3$, where $U(1)_{\mathrm{down}}$ is the linear combination of U(1)$_r$ 
and U(1)$_t$ that gives a neutral right-handed down quark. To be more precise,
if we choose the generator of 
$U(1)_{\mathrm{down}}$ to be the linear combination   $Y_r-3 Y_t$
 (see Table \ref{tab1}), then only the following operators are $G^\prime$
 invariant  
\begin{equation}
g \bar d_L^i d_R^i H^i \phi_i + g_L \bar d_L^i d_R^j H^i \phi_i
\end{equation}
In fact the fermionic field $d_R^j$ does not carry charge with respect 
 $U(1)_{\mathrm{down}}$, whatever it is $j$.  Instead the group  symmetry
  $G^\prime=(U(1)_{\mathrm{down}})^3 >{\hspace{-6pt}}\triangleleft  S_3$
 forbids 
the operators with couplings  $g_R$ and $g_3$. It is 
at the breaking scale of $G^\prime$ that also these operators appear.
\subsection*{Charged lepton masses}
Now let us try to extend the discussion to charged leptons and neutrinos.
In $E_6$, the charged leptons have the same $U(1)$ charges of down quarks,
once  we exchange the left-handed components with the right-handed ones.
In fact the   $d_L$ has the same charges of the lepton 
 $e_R^c$, while  $d_R^c$ has the same charges of the lepton $e_L$ 
(see Table 1). 
In principle , we could expect that the charged lepton mass matrix is
similar in terms of order of magnitude  to the down quark mass matrix,
except for a transposition operation that exchanges left-handed components 
with right-handed ones.
 In SU(5) unified models,  often it happens that the down quark matrix 
is exactly the transposed of the charged lepton matrix. 
If we apply such a transposition operation to the down quark mass matrix 
 obtained in the model of  Table \ref{tab2}, and we take it as the charged 
lepton mass matrix, we obtain that the unitary matrix that diagonalizes it 
 is very similar to the   $V_{\mathrm{pmns}}$ mixing matrix that fit all 
neutrino  oscillation data.
This result would be very interesting, because it would provide us 
with an explanation of neutrino oscillation data and their peculiarities,
 with no further effort. A similar matrix has been discussed in 
\cite{Fritzsch:2004xc}.
However such  a scheme of exact SU(5) unification does not work.
In fact lepton physical masses are different from the down quark masses at
 the   unification scale, in particular the electron is much lighter 
than the down quark, and this contradicts the fact that the transposed 
matrix has the same eigenvalues of the original one.
The minimal SU(5) with  the exact unification must be discarded, but  
a relation between the elements of the two matrices 
(lepton and down quark ones) in terms only 
of order of magnitudes is still allowed. Namely the relations below
\begin{eqnarray}
g^d_L&\simeq g^l_R \nonumber\\
g^d_R&\simeq g^l_L \nonumber\\
g^d_3&\simeq g^l_3 \label{gql1}
\end{eqnarray}
are compatible with the charges of the group G.
If relations (\ref{gql1})  would be exact, we would be back to 
the exact minimal SU(5) unification, and the electron mass would be too 
large\footnote{Another  possibility, not explored here, is to add an 
abelian factor $(U(1)_{\mathrm{hyper}})^3$ to the group G, whose generator 
is proportional to the usual hypercharge. Such an extended G group would
 contain charges that distinguish leptons and down quarks and would not 
give relations (\ref{gql1}).}.
To reconcile the electron mass with  its experimental value is enough 
to increase $g^l$ by a factor of two, with respect $g_d$ 
(see Table \ref{tab3}), and 
this is not  incompatible with the group G.
The increase of  $g^l$ improves the fit of all charged lepton physical 
 masses, but 
 the mass  matrix becomes closer to a diagonal matrix; as a consequence 
the oscillation angles deduced from it are no longer large enough 
to explain flavor oscillation in neutrino phenomenology.  

Namely taking  the $g$ couplings for the charged leptons as in Table 
\ref{tab3}
\begin{table}[h]
$$
\begin{tabular}{ll}
\hline
down&lepton\\
\hline
&\\
$g^d=0.848$ & $g^l=1.747$ \\
$g^d_L=0.305~e^{-0.02~i}$&$g^l_L=0.0048~e^{-0.12~i}$\\
$g^d_R=0.002~e^{-0.12~i}$&$g^l_R=0.152~e^{-0.02~i}$\\
$g^d_3~=0.009~e^{1.82~i}$&$g^l_3~= 0.0085~e^{1.82~i}$\\
\hline
\end{tabular}
$$
\caption{$g$ couplings in the down quark sector and charged lepton sector.
The mass scale has been set as in Table 2.}\label{tab3}
\end{table}
 and the following lagrangian 
\begin{equation}
\sum_{i,j,k} \left(g^l \bar e_L^i e_R^i H \phi^d_i + g_L \bar e_L^i e_R^j H 
\phi^d_i+
g^l_R \bar e_L^i e_R^j H \phi^d_j+g^l_3 \bar e_L^i e_R^j H \phi^d_k\right)
\end{equation}
we get the following mixing angles
\begin{equation}
V_l=\left(
\begin{array}{ccc}
0.991\,e^{3.029\,i } & 0.100\,
   e^{-1.131\,i } & 0.082 \\
 0.096\,   e^{-2.161\,i } & 0.991\,e^{3.035\,i } &
 0.091 \\
 0.087\,e^{-0.021\,i } & 0.087\,
   e^{-0.025\,i } & 0.992    
\end{array}   \label{vl}  
\right).
\end{equation}
The explanation for   neutrino oscillations will be  given in 
the next section in terms of new symmetries in the neutrino sector.  
\subsection*{Neutrino oscillations}
In the previous section we have seen that extending the standard model 
with an additional gauge symmetry group
 $ G=(U(1)_r\times U(1)_t)^3 >{\hspace{-6pt}}\triangleleft S_3$
 we are able to give a realistic explanation for masses and mixings in 
the quark sector. This is also compatible with the physical charged lepton 
masses.
The $G$ group can be embedded in a larger and unified symmetry 
group. Furthermore it is not in contradiction with physical quark 
and lepton masses, differently from  the minimal SU(5) 
predictions\footnote{Minimal SU(5) 
exact unification is also disfavored by the standard model measurement 
of gauge coupling that do  not exactly unify at the unification scale
\cite{Altarelli:2000fu}. }.
 
After the assumption of a family  permutation symmetry,
 as derived from third quantization, we have been lead 
to add new abelian group factors to the gauge group  in order to explain 
quark mass and  mixings, and the right hierarchies among different mass
 operators.
   
Let us try to explain large neutrino mixings.
In a previous work \cite{Caravaglios:2005gw} we have shown that the $S_3$
 symmetry and its breaking into 
$S_2$, can explain the peculiar pattern of the  $V_{\mathrm{pmns}}$.
In fact, in that paper we started from the matrix below
\begin{equation}
m_{\mathrm{maj}}= \left( 
\begin{array}{ccc}
a+b+\varepsilon & a & a \\ 
a &a+ b & a \\ 
a & a &a+ b
\end{array}
\right)  \label{10}
\end{equation}
showing that it is diagonalized by the so-called 
tri-bimaximal matrix in the limit $ a \gg \varepsilon$ 
\begin{equation}
V_{\nu}=\left( 
\begin{tabular}{lll}
$\frac{-2}{\sqrt{6}}$ & $\frac{1}{\sqrt{3}}$ & $0$ \\ 
$\frac{1}{\sqrt{6}}$ & $\frac{1}{\sqrt{3}}$ & $\frac{-1}{\sqrt{2}}$ \\ 
$\frac{1}{\sqrt{6}}$ & $\frac{1}{\sqrt{3}}$ & $\frac{1}{\sqrt{2}}$%
\end{tabular}
\right) .\label{Op}  
\end{equation}
The matrix (\ref{10}) is $S_3$ symmetric in the limit  $\varepsilon \ll a,b$,
 while 
the small parameter  $\varepsilon$ breaks $S_3$ into $S_2$.
From one side the matrix (\ref{10})  gives a simple explanation 
for the mixing angles in the  $V_{\mathrm{pmns}}$, on the other side 
 it does not give a clear motivation for 
the mass hierarchies in  solar and atmospheric experiments. Additional
 hypothesis are needed to justify  the right hierarchy.
In fact the mass matrix (\ref{10}) has two almost degenerate mass eigenstates, 
(if
 $\varepsilon\ll a,b$) both with mass $b$ and one state with mass $3 a +b$.
Neutrino phenomenology requires that the non degenerate state, that usually 
is the heaviest one, does not mix with the electron neutrino, and it is 
maximally mixed with the tau and muon neutrino. 
This is not consistent with the diagonalization of the matrix (\ref{10}), 
where 
the non degenerate state, with mass $3 a + b $ is the $S_3$ singlet, thus 
it has a large mixing with the electron neutrino.
Further effort is needed to understand the solar and atmospheric masses.
 We have to specify the values of  $b,a,\varepsilon$
and understand how to 
modify their  relations with solar/atmospheric mass data.
It is easy to realize that we have to take  $b\ll\varepsilon \ll a$
in order to have three masses with  different order of magnitudes. But 
this is not yet enough, because the lightest neutrino is the one with mass $b$,
that does not mix with the electron neutrino, while it must be the heaviest
in order  to be compatible with the experiments.
To achieve the final result, we also need to make neutrino 
 masses proportional to the inverse of the free parameters
  $b,\varepsilon,a$.
Therefore we exploit the seesaw mechanism as follows.
Let us imagine that left-handed  neutrino masses comes out 
from the usual seesaw mechanism.
In this case the light neutrino mass matrix is proportional to 
\begin{equation}\label{seesaw}
M_L=M_{\mathrm{dirac}}~M_{\mathrm{maj}}^{-1}~M_{\mathrm{dirac}}^t
\end{equation}
where  $~M_{\mathrm{maj}}$ is the majorana mass of the right-handed neutrino 
while  $M_{\mathrm{dirac}}$ is the dirac mass between the left-handed and
 right-handed neutrinos.
In  eq. (\ref{seesaw})  the majorana masses $~M_{\mathrm{maj}}$ are 
typically of the
 order of the unification scale, while  $M_{\mathrm{dirac}}$ is of the 
order of the electroweak scale.
If  $M_{\mathrm{dirac}}$ is proportional to the identity matrix, the unitary 
matrix that diagonalizes  $M_L$ is identical to the one that diagonalizes
 $~M_{\mathrm{maj}}$, while the eigenvalues of  $M_L$ will be proportional 
to the inverse of the eigenvalues of $~M_{\mathrm{maj}}$.
Namely, if  $b\ll \varepsilon\ll a$, if $~M_{\mathrm{maj}}$ is equal to 
eq. (\ref{10})
and if 
\begin{equation}\label{mD}
M_{\mathrm{dirac}}=m I 
\end{equation}
where $I$ is the identity matrix, we get 
the following physical masses for the lightest neutrinos
\begin{eqnarray}
m_1&= \frac{3 m^2}{2 \varepsilon}\nonumber\\
m_2&= \frac{m^2}{3 a}\nonumber\\
m_3&= \frac{m^2}{b}
 \end{eqnarray}
while the diagonalization unitary matrix is the (\ref{Op}).
This result is known  \cite{Caravaglios:2005gw}, 
but now we want to deduce the matrices (\ref{10}),(\ref{mD})
  and the hierarchy $a \gg\varepsilon \gg b$ from 
new fundamental symmetries and their spontaneous breaking.
Explaining the matrix (\ref{mD}) is easy, 
since it is enough to extend 
the previously discussed symmetries 
 in the quark sector (see the previous section) also  to neutrinos:
 observing that the Higgs $H_i$ of the $i$-th family has charges $(q_r,q_t)
= (+3,+5)$
with respect the U(1) of the $i$-th family. In this case 
the following Yukawa interaction 
\begin{equation}\label{Yg}
\sum_{i} g \bar\nu_L^i X_L^i H^i
\end{equation}
is the only Yukawa coupling allowed by the $G$ symmetry. In fact the fermion 
 $X_L^i$ is the singlet of  SU(3)$\times$SU(2)$\times$U(1)
 with $(q_r,q_t)= (0,4)$
with respect the additional  U(1) of the $i$-th family (see Table 1).
It is easy to check that  $ g \bar\nu_L^i X_L^j H_i $ is  
$S_3$ invariant but for 
 $i\ne j$ the   U(1)-charges  of the $i$-th family are not conserved.
Note that the G symmetry allows for the tree level coupling (\ref{Yg}) 
only for neutrinos.
Quark and charged leptons 
(including up quarks that usually unify with neutrinos in SO(10))
need higher mass dimension operators, that is operators including a new field 
 $\phi_i$ with non zero U(1)  charges to make a gauge invariant term.
This is an important property of  the model (see \cite{Caravaglios:2005gf}).
This is 
 to avoid 
a unification relation between the up quarks and neutrinos, that 
 would make difficult to make consistent  mass and mixings in the up quark 
sector with those observed in the neutrino sector\footnote{To avoid 
the unification between up quarks and neutrinos we have added two U(1) of 
$E_6$. The unique  U(1) that comes from SO(10) would not be sufficient
\cite{Caravaglios:2005gf}.}.
It is not difficult to reproduce the Dirac mass matrix of eq.(\ref{mD}),
 but the 
matrix of eq. (\ref{10}) and in particular the hierarchy  
 $b\ll \varepsilon\ll a$, cannot 
be deduced in an obvious way from the model with the $G$ symmetry that we have 
considered above.
In fact it is not possible to deduce from the $G$ symmetry that 
  the majorana masses below 
\begin{equation}\label{14}
\sum_{ij}\left( m_0 X_L^i X_L^i+m_1   X_L^i X_L^j\right)
\end{equation}
 satisfy  $m_0\ll m_1$ or equivalently $b \ll a$. 
Unfortunately the  $G$ symmetry  implies    $m_0>m_1$.
We can try to enlarge the flavor symmetry $G$ in order to understand this
difference. Third quantization requires  that the flavor symmetry is 
the permutation symmetry of all families \cite{Caravaglios:2002ws}. 
Thus we can enlarge $S_3$ into $S_4$ , but we have to introduce a fourth 
family.
This fourth family cannot have the same quantum numbers of the first three
families with respect the standard model gauge group, otherwise,  
 we would have seen it.\footnote{Precision tests and direct searches seems 
to disfavor this possibility. Also the number of light neutrinos measured 
at LEP, appears to forbid this possibility.}
Third quantization and $S_4$ symmetry offer  the possibility 
to extend the concept of family to all gauge bosons  (and not only 
to the U(1) as we have done in the previous section), and to avoid 
the problem of having four chiral families at the electroweak scale.
We will show in the following how to proceed.
In the previous section we have only mentioned the possibility 
to consider a grand unified group, we are now able to give an explicit 
choice. Let us consider the group 
E$_6^4 >{\hspace{-6pt}}\triangleleft$ S$_4$. 
We remind that  $>{\hspace{-6pt}}\triangleleft$ means the
 semidirect product: the action of the elements 
of $S_4$, onto the gauge group  $E_6^4$ is to permutate the different 
 $E_6$  factors  of the full group  $E_6^4$. In other 
words the concept of family is extended  to the gauge group and the 
 gauge bosons. In this specific case,  before the spontaneous symmetry
breaking,  each  family contains 27 fermions (the smallest irreducible 
representation of $E_6$) and 78 gauge bosons ($E_6$ gauge bosons).
Fermions and bosons of each family interact among them, but do not interact 
with fermion and bosons of a distinct family.
After the spontaneous symmetry breaking of the gauge symmetry, the standard 
model 
gauge bosons are a linear combination of the gauge bosons of the first 
three families. This explains why only the first three families have non zero
 charges, with respect  SU(3)$\times$SU(2)$\times$U(1) of the standard model. 
For example, if we label  $\lambda^a_i$ the generator of the color  $SU(3)$ 
of the $i$-th family ({\it i.e.} it  acts only on the fermions of the 
 $i$-th family) we have that the generator of the QCD group of the standard 
model, will be given by the linear combination
 \begin{equation}\label{lamb}
\lambda^a=\lambda^a_1+\lambda^a_2+\lambda^a_3.
\end{equation}
In this case the standard model gluon $A^\mu_a$, that is contracted 
 to the generator $\lambda^a$,  only interacts with the first three families, 
the generator $\lambda^a_4$ being missing in the combination (\ref{lamb}).
It is not difficult to imagine a spontaneous symmetry breaking pattern of the 
$E_6^4$  symmetry that gives the generators in eq. (\ref{lamb}) as the only 
unbroken 
generators of the group SU(3). One can easily check that the group 
$G$ of the previous section is a subgroup of  
E$_6^4 >{\hspace{-6pt}}\triangleleft$ S$_4$, and we understand that 
 the group $G$ appears at 
an intermediate scale,  as the remaining 
unbroken symmetry after spontaneous breaking of the full group.

We now proceed to see  what happens  in the neutrino sector and why such a 
large unified group helps in understanding the mass hierarchy of the 
 singlet/doublet $S_3$ components  ($m_0\ll m_1$ in eq. (\ref{14}))
 and the large splitting between the doublet 
components. 
Let us write the most general renormalizable lagrangian, compatible 
with 
 E$_6^4 >{\hspace{-6pt}}\triangleleft$ S$_4$, focusing  
on the neutrino sector (the indices $ij$ now run from 1 to 4)
\begin{equation}\label{15}
\sum_{i,j} \left(g \nu_L^i X_L^i H^i+ g_1 \nu_R^i X_L^j \omega^{ij}\right). 
\end{equation}
We have added a scalar field  $\omega^{ij}$, that is a singlet of the standard
model. It takes a vev, giving a Dirac mass that mixes   $\nu_R$ and  $X_L$.
The reason of this choice is the following.
We have chosen a tensor with two indices $ij$, because a field 
 $ \alpha^{i} $ with only one index would introduce a diagonal mass 
term  $ g_1 \nu_R^i X_L^i \alpha^{i}$. This term gives  mass both to the 
doublet and the singlet of $S_3$, mixing them.
Our aim is different, we want to give mass only to the singlet, leaving 
the doublets massless.  This hierarchy between singlets and doublets
is necessary, as we have explained  after the example of eq. (\ref{14}). 
After having discarded the tensor $\alpha^i$, we can 
 consider the two indices 
 tensor  $ \omega^{ij} $: if we take it as a symmetric tensor under the 
exchange of the $i$ and $j$ indices, then this scalar field could take a vev 
in the direction   $ \omega^{ii}=v $ and we obtain  a mass both for the 
singlet  and the doublets. There is only one peculiar direction 
   $ \omega^{ij}=v $ (for any choice of $i$ and $j$) that gives 
the hierarchy that we need, but we have no 
fundamental  argument to justify this choice.

 Instead, if we take  an antisymmetric tensor  $ \omega^{ij}$  such that 
$ \omega^{ij} =- \omega^{ji}$, then  the scalar field $\omega$ no longer has 
diagonal elements  ($ \omega^{ii}=0$, by definition of antisymmetric tensor)
and as we will see it will give mass only to the singlets of $S_3$.
A comment is in order here, on the choice of the scalar field  $ \omega^{ij}$:
third quantization predicts that if a field belongs to a family 
it must carry  only one index $i$. It seems that  the scalar field 
 $ \omega^{ij}$  does not satisfy this rule. However 
we can imagine that the scalar field  $\omega^{ij}$  is not a fundamental 
field, but it could be a fermion condensate 
$\left<\omega^{ij}\right>=\left< \eta^i \psi^j\right>$. 
In this case the fundamental fields  $\eta^i$ and  $\psi^j$ 
carry only one family index as predicted by third quantization.
Being an antisymmetric tensor, the scalar field  $ \omega^{ij}$ can 
 induce  only one renormalizable yukawa interaction, that is the one in
 eq. (\ref{15}).
It mixes  $\nu_R$ with  $X_L$, and it cannot mix fields of the same 
kind ( $X_L^i X_L^j \omega^{ij}=0$ for the antisymmetric nature 
of $\omega$).
We properly choose the charges of the scalar field $\omega^{ij}$ in order 
to make the lagrangian (\ref{15})   E$_6^4 >{\hspace{-6pt}}\triangleleft$ S$_4$
 invariant.
The antisymmetric field $\omega^{ij}$ does not contains $S_4$ singlet and 
it must break it. It contains only one $S_3$ singlet, so it can break 
$S_4$ into $S_3$ as we desired. 
Namely, if $S_4$ is broken and the $S_3$ is unbroken the scalar field
$\omega^{ij}$  can take
a vev only in one direction   
\begin{equation}
\left<\omega^{ij}\right>= \left( 
\begin{array}{cccc}
0 & 0 & 0 & v \\ 
0 & 0 & 0 & v \\ 
0 & 0 & 0 & v \\ 
-v & -v & -v & 0 \\ 
\end{array}
\right)_{ij}. \label{13}
\end{equation}
It is easy to check that the permutation of the first three families 
leaves the vev (\ref{13}) invariant. For completeness, we give an example of 
scalar potential that can give the desired vev  as  in eq. (\ref{13})
\begin{eqnarray}
V=&-\mu_1^2 \sum_i h_i^2+\mu_2^2 \sum_{ij}|\omega_{ij}|^2+ \lambda_1 
\sum_{ij}|\omega_{ij}|^4+ \lambda_2 \left(\sum_{ij}|\omega_{ij}|^2\right)^2+
\\ \nonumber
+&  \lambda_3 \sum_{ijkl} \omega_{ij}\omega^{lj*}\omega_{lk}\omega^{ik*}-
\lambda_4  \sum_{ij}|\omega_{ij}|^2 h_j^2 -\lambda_5 \sum_i h_i^4
+\lambda_6 \left(\sum_i h_i^2 \right)^2. 
\end{eqnarray}
Note that we have added a scalar field $h_i$ that is a singlet under 
$E_6^4$, and transform as a triplet  ({\bf $3_1$}) under S$_4$. This 
additional field seems to be necessary if we want that the vev (\ref{13}) is 
a minimum of the potential  and if we want the breaking  
$U(1)_{\mathrm{down}}^4 >{\hspace{-6pt}}\triangleleft$ S$_4$ $\supset S_3$.
Putting this vev in eq. (\ref{15}), we obtain the mass for 
some neutrino components 
\begin{equation}
g \nu_L^i X_L^i H^i+ g_1 v \nu_R^i X_L^4  - g_1 v \nu_R^4 X_L^j
+{\mathrm{h.c.}} 
\end{equation}
We get two Dirac neutrinos  $(\nu_R^4,X_L^s)$ and   $(\nu_R^s,X_L^4)$ 
with mass at the breaking scale of $S_4$, $ g_1 v$.
These two Dirac fermions are very much heavy and are composed by 
four weyl fermions  $\nu_R^4$, 
$X_L^4$, $X_L^s=X_L^1+X_L^2+X_L^3$,  $\nu_R^s=\nu_R^1+\nu_R^2+\nu_R^3$, that 
transform as singlets under $S_3$: all doublets of $S_3$ remain
massless.
This accomplishes our task, the same configuration given in the example 
 eq. (\ref{10}-\ref{14}), 
is achieved here: very heavy $S_3$ singlets compared to 
 very light doublets\footnote{All these neutrinos are expected to be much 
heavier than the weak scale and close to the unification scale.}. 
The diagonalization of the neutrino mass matrix\footnote{See the appendix B.}
 proceeds  
 similarly to the diagonalization of the simplified model of eq. (\ref{10}):
 the main difference  is that in the simplified model eq. (\ref{10}) 
we only 
have a majorana mass for the three majorana fermions $X_L^i$. In this  model 
the dominant mass term 
 mixes  $X_L^i$ and $\nu_R^i$, and  forms very heavy dirac fermions. In both 
cases the unitary matrix that diagonalizes the light left-handed neutrinos
is given by the tri-bimaximal matrix (\ref{Op}). Deriving the mass eigenvalues 
 is less trivial than the usual seesaw. 
First we have  to give  mass  to the $S_3$ doublets that are combinations of 
 $X_L^i$ and
 $\nu_R^i$. 
We introduce two  scalar fields  $\xi^i$ and $\chi^i$ 
(since we have enlarged the permutation group, these are 
quadruplets under $S_4$).
 They break $S_3$ into $S_2$  through the vev 
  $\xi^1=v_1$, $\xi^2=\xi^3=v_2$, $\xi^4=v_4$  and 
 $\chi^1=v^\prime_1$, $\chi^2=\chi^3=v^\prime_2$,
$ \chi^4=v^\prime_4$ with  $v_2\ll v_1\ll v$ and 
 $v^\prime_2\ll
 v^\prime_1\ll v$.
 The full lagrangian in the neutrino sector now becomes 
 \begin{equation}
g \nu_L^i X_L^i H^i+ g_1 v \sum_{i=1}^3 \nu_R^i X_L^4  - 
g_1 v \sum_{j=1}^3 \nu_R^4 X_L^j +
 g_2  \sum_{i=1}^4 X_L^i X_L^i \xi^i+ g_3 \sum_{i=1}^4  \nu_R^i 
\nu_R^i {\chi^i} +{\mathrm {h.c.}} \label{eq.18}
\end{equation}
As shown in the appendix, we obtain the following masses and mixings for the 
 light left-handed neutrinos when also the Higgs  
field H takes a vev \footnote{The component $H_4$
of the field is a standard model singlet and it takes a vev much larger 
than the electroweak scale, while all the first three components 
of $H_i$ take the same vev $H$. This is because only the $S_3$ 
 singlet component of $H^i$ takes a vev. 
For an explanation, see the next section.}   
$\langle H_i\rangle=(H,H,H,H_4)$ 
\begin{eqnarray}
m_1 &=\frac{3\,g^2\,H^2}
  {2\,g_2\,{v_1} + 
    {g_2}\,{v_2}} \\
m_2 &=\frac{g^2\,H^2\,{v_4^\prime}}{3\,{{g_1}}^2\,v^2} \\
m_3 &=\frac{g^2\,H^2}{{g_2}\,{v_2}}. 
\end{eqnarray}
These give the following solar and atmospheric masses in the limit\footnote{
This hierararchy seems in contradiction with that one observed in the 
charged lepton sector $m_\tau\gg m_e$,also shown in Table 2. 
See  the next section for an explanation. We obtain a negative 
$\Delta m^2_{\mathrm{sol}}$, while solar neutrino data  prefer a positive 
number, see ref.\cite{Fogli:2005cq} for the data analysis. 
Our  prediction can only  explain
$|\Delta m^2_{\mathrm{atm}}|\gg| 
\Delta m^2_{\mathrm{sol}}|$. In order  to predict the correct sign, further 
investigation is needed.} 
 $  v \gg  v_4^\prime \gg v_1 \gg v_2$
\begin{eqnarray} \label{predizioni}
\Delta m^2_{\mathrm{atm}}&=\frac{g^4\,H^4}{{g_2}^2\,{v_2}^2} \nonumber\\
\Delta m^2_{\mathrm{sol}}&=-\frac{9\,g^4\,H^4}
  {4\,g_2^2\,{v_1^2}} 
 \end{eqnarray}
and the following mixing matrix 
\begin{equation}
V_{\nu}=\left( 
\begin{tabular}{lll}
$\frac{-2}{\sqrt{6}}$ & $\frac{1}{\sqrt{3}}$ & $0$ \\ 
$\frac{1}{\sqrt{6}}$ & $\frac{1}{\sqrt{3}}$ & $\frac{-1}{\sqrt{2}}$ \\ 
$\frac{1}{\sqrt{6}}$ & $\frac{1}{\sqrt{3}}$ & $\frac{1}{\sqrt{2}}$%
\end{tabular}
\right) .  \label{1}
\end{equation}
Combining the $V_l$ obtained in the previous section, eq. (\ref{vl})
 and  the $V_{\nu}$
we get the following 
 predictions for the solar and atmospheric   mixing  angles
\begin{eqnarray*}
&&\sin^2(\theta_{12})=0.25 \\
&&\sin^2(\theta_{23})=0.58\\
&&\sin^2(\theta_{13})=0.005.
\end{eqnarray*}
These mixing angles and the hierarchy $|\Delta m^2_{\mathrm{atm}}|\gg| 
\Delta m^2_{\mathrm{sol}}|$ (see eq.(\ref{predizioni})) 
 are in  agreement with the experimental
 observation, but the sign of $\Delta m^2_{\mathrm{sol}}$ is wrong. Further 
study is needed to predict the correct sign.

Let us summarize the model.
We start with the group  E$_6^4 >{\hspace{-6pt}}\triangleleft$ S$_4$.
This group is broken into 
 $(SU(3)\times SU(2) \times U(1) \times U(1)_r \times U(1)_t)^4
  >{\hspace{-6pt}}\triangleleft$ S$_4$ at the intermediate  scale  $M_X$.
The  $(U(1)_r \times U(1)_t)^4$ factor breaks  at the scale $M_X^\prime$
and we remain with the 
 $(SU(3)\times SU(2) \times U(1) \times U(1)_{\mathrm{down}})^4
  >{\hspace{-6pt}}\triangleleft$ S$_4$
symmetry.  Where  $ U(1)_{\mathrm{down}}$ is the combination of  $U(1)_r$ and
 $ U(1)_t$  that leaves both the right-handed quark and the left-handed
 charged leptons neutral.
There is another scale   $M_X^{\prime\prime}$ below the  $M_X^\prime$
at which the group is further broken into 
 $SU(3)\times SU(2) \times U(1)\times $ S$_3$. 
If $\Lambda$ is the characteristic scale of the fundamental theory 
 E$_6^4 >{\hspace{-6pt}}\triangleleft$ S$_4$
 we could expect the following hierarchies for all couplings
\begin{eqnarray}
g_L^d\simeq \frac{M_X^{\prime}}{\Lambda^2}; &
g_R^l\simeq \frac{M_X^{\prime}}{\Lambda^2} \nonumber \\
g_R^d\simeq \frac{M_X^{\prime\prime}}{\Lambda^2}; &
g_L^l\simeq \frac{M_X^{\prime\prime}}{\Lambda^2}\nonumber \\
g_3^d\simeq \frac{M_X^{\prime\prime}}{\Lambda^2}; &
g_3^l\simeq \frac{M_X^{\prime\prime}}{\Lambda^2}\nonumber \\
g_R^u\simeq \frac{M_X^{\prime\prime}}{\Lambda^2}; &
g_L^u\simeq \frac{M_X^{\prime\prime}}{\Lambda^2}\nonumber \\
g_3^u\simeq \frac{M_X^{\prime\prime}}{\Lambda^2}; &
v_2\ll v_1\ll M_X^{\prime\prime} \nonumber
\end{eqnarray}
that are in agreement with the values derived from the experimental data
(see Table 2 and 3)). 
\subsection*{Neutrino and charged fermions  misalignment in the $S_3$ symmetry 
breaking}
In the previous sections we have discussed the $S_3$ 
 spontaneous symmetry breaking  and its implications to the fermion 
mass matrices. To each fermion type, we have associated its own 
scalar $S_3$ triplet, responsible for the $S_3$ breaking. 
We have  associated the scalar field $\phi_i^u$, to the up quark sector.
 The down 
quark and charged lepton mass matrices appear after the field  
$\phi_i^d$ take a vev. Finally the field $\xi_i$ and $\chi_i$ are responsible
 for the $S_3$ breaking in the neutrino sector.
We have found 
that differently from the charged fermions, where the hierarchy proceeds 
in the direction $\phi_3\gg \phi_2\gg \phi_1$, 
we need an inverted hierarchy for the scalar fields   
 $\xi_i$ and $\chi_i$, namely  
  $\xi_1\gg \xi_2\simeq \xi_3$. An explanation of this special feature requires
a detailed analysis of the effective   scalar potential of the full lagrangian.
We have not done this study. Instead we will show  a rather 
 simple  mechanism that could give the desired breaking and  we postpone 
a more rigorous proof in a future work. 
Let us assume that only two scalar fields $\xi_i$ and $\phi_i^u$ are 
responsible
for the $S_3$ breaking and that the other fields $\phi_i^d$ and  $\chi_i$
are not fundamental fields but just products and combinations of other 
fondamental  fields. 
The full effective scalar potential is 
\begin{equation}
V=V_1(\xi_i)+V_2(\phi_i^u)+V_3(\phi_i^u,\,\xi_i)\label{pot}
\end{equation}
and it is invariant under  $(U(1)_r \times U(1)_t)^4
 >{\hspace{-6pt}}\triangleleft S_4$. The charges of $\xi_i$ and $\phi_i^u$ 
can be deduced from eq.(\ref{up},\ref{eq.18}) and Table 1.
 It is not difficult to show that
 gauge invariance implies  both fields $\xi_i$ and  $\phi_i^u$ must appear
 as  $2 n$  powers (with $n$ integer) in the potential (\ref{pot}). 
If all mixed terms like $\lambda \sum_i |\xi_i \phi^u_i|^2$  have a positive 
$\lambda$ coefficient   then the fields  $\xi_i$ and  $\phi_i^u$  have 
orthogonal vev  (misaligned). 
This is precisely what is  required by the model discussed in the previous 
sections. 
 
\subsection*{The electroweak symmetry breaking and the flavor symmetry $S_3$}
In the previous sections we have seen that the Higgs doublet responsible for
 the electroweak symmetry breaking has to be chosen in the  triplet reducible 
representation of $S_3$ in order to make the lagrangian (3) gauge invariant 
under  $(U(1)_{\mathrm{down}})^3 >{\hspace{-6pt}}\triangleleft S_3$.
At the same time we want that there is only 
one Higgs doublet at the electroweak scale. 
This is to avoid dangerous flavor changing 
neutral currents due to the virtual  exchange of additional Higgs fields.
In this section we will see that requiring  just  one Higgs doublet  at 
the weak scale implies that only one component of the original 
triplet $H^i$ (with $i=1,3$) can take a vev, and this component must be a 
singlet of $S_3$. This is true even if the breaking of $S_3$ occurs at a
 much higher scale.
The reason is the following.

We can split the effective potential of the three Higgs doublets $H^i$ 
as follows 
\begin{equation}
V(H^i)=V_{S_3}(H^i) +\Delta V(H^i,\phi^i)
\end{equation}
where  $V_{S_3}$ is an invariant potential under $S_3$ while  $\Delta V$ 
 breaks $S_3$ because it contains all scalar fields  $\phi^i$ that break 
the $S_3$ symmetry. $\Delta V$ is 
defined in such a way that 
 $\Delta V(H^i,\phi^i)=0$ in the limit  $\phi^i=0$. Splitting the potential 
in a $S_3$ symmetric part plus a $S_3$ not symmetric part is always possible. 
The Higgs mass matrix $M_{kj}$,  ($k,j$ are family indices)    is  given by 
\begin{equation}
M_{kj}=\frac{\partial^2}{\partial H_k \partial H_j} V(H_i).
\end{equation}
If we only consider terms coming from $V_{S_3}(H^i)$, the mass matrix 
$M_{kj}$ will be $S_3$ invariant. Thus we expect  two degenerate eigenstates
with mass $m_D$ that transform as a doublet of $S_3$,  plus one 
eigenstate with mass   $m_S$ that is a $S_3$ singlet.
We remind that  we require just one light electroweak 
Higgs. With this assumption
 we must have    $|m_S^2|<<|m_D^2|$. Furthermore only the $S_3$ 
singlet component of $H^i$ can take a vev.
In fact a possible vev of the doublet component would be proportional to 
$m_D$. But, at the same time, we know that  $|m_D^2| \gg M_W^2$, and therefore 
the doublet cannot take a vev.

Now it is easy to understand that, since  $ \Delta V(H^i,\phi^i)$ is just a 
perturbation of the $S_3$ symmetric part,  the 
minimization of the full potential  $V(H_i)$ can only give a very tiny 
mixing between the doublet and singlet components of $H^i$, 
being\footnote{ $m_D$ is expected to be not much smaller than the unification 
scale while  $\phi^i$  is the breaking scale of $S_3$.}
 $m_D>>\phi^i$. 
This argument does not show in a rigorously way that the Higgs doublet 
must be the $S_3$ singlet, but show that is a realistic and reasonable
hypothesis \cite{Kubo:2004ps}.
 \subsection*{Conclusions}
The experimental mass measurements in neutrino oscillations, together 
with  masses and mixings of the rest of matter fermions, give us an almost 
complete picture of the less understood sector of the standard model, that 
is the sector of Yukawa interactions. We also know that 
 grand unification can explain charge quantization, and can 
 embed all fermions of each family in just one  irreducible 
representation of SO(10). 
The simplest and minimal SO(10) model predicts that the up quark 
mass matrix is equal to the dirac neutrino matrix.
Minimal SU(5) predicts that the mass matrix of charged leptons is the 
transposed of the down quark mass matrix. Both predictions are not 
in agreement with the experimental observation. An effort to 
go beyond the minimal unification scenario is needed.
We also know, that if the fundamental lagrangian of interactions 
comes from the third quantization, the flavor symmetry group 
is very likely to be the permutation symmetry. This symmetry group 
gives good predictions of neutrino oscillations. 
 How to extend such a symmetry to the quark sector 
 in the context of a grand unified group is the subject of this work.
In particular we improve some aspects of the model studied in 
\cite{Caravaglios:2005gw,Caravaglios:2005gf}
 clarifying why in eq.(\ref{10})   $a \gg b$. 
We have found that the unification group 
E$_6^4 >{\hspace{-6pt}}\triangleleft$ S$_4$, with $S_4$ the flavor group and 
$E_6^4$ the gauge group, predicts at least one breaking pattern 
that simultaneously explain  masses and mixings of quarks, leptons
 and neutrinos.
It is  important  to note that  the symmetry group  is the semidirect 
product   (and not the direct product) between the gauge group and the 
flavor group.
This can descend in a natural way from third quantization and predicts 
why operators (or interactions) that mix different families 
(see the discussion after eq.(\ref{down})) are smaller than  those that do 
not mix 
families.
We have put in evidence that 
$U(1)_{\mathrm{down}}^4 >{\hspace{-6pt}}\triangleleft$ S$_4$,
that is a subgroup of E$_6^4 >{\hspace{-6pt}}\triangleleft$ S$_4$,
predicts the right hierarchies among the couplings of 
 the operators needed to explain 
quark mass and mixings.
We have found that the introduction of the  scalar 
field   $\omega_{ij}$,
antisymmetric with respect $i$ and $j$  is useful. Because it explains 
why the $S_3$ singlet components of $X_L$ and $\nu_R$ are much heavier 
than the doublets. As a consequence  we also get 
$\Delta M_{\mathrm{atm}}^2 \gg 
  \Delta M_{\mathrm{sol}}^2$.
This hierarchy  was assumed as a starting hypothesis in 
\cite{Caravaglios:2005gw}, now a 
clear motivation for this choice has been given. 

Finally we know that  the possibility 
of exact SU(5) unification in the yukawa sector is excluded, but
 we have shown that partial unification, {\it i.e. } in terms of order 
of magnitudes is still possible. Namely, 
in our model it exists an intermediate scale, at which 
the U(1) factors contained in  $E_6^4$ are unbroken. This abelian group 
implies that each mass  operator  of the down quarks is of the same order
of magnitude of its analogue  in the charged lepton, taking into account 
that right-handed quarks are partners of left-handed charged leptons.
In particular 
we have found that a  factor two in the relative 
size  between quarks and lepton operators is enough to obtain the electron 
and down 
quark mass ratio, and neutrino oscillations  in agreement with data.

To understand these small differences in the relative size
between  quark/lepton operators
 we need a deeper analysis of the symmetry breaking  pattern of  
E$_6^4 >{\hspace{-6pt}}\triangleleft$ S$_4$.

\subsection*{Appendix A. E$_6^4 >{\hspace{-8pt}}\triangleleft$ S$_4$ 
irreducible   representations}
The permutation  group  $S_4$ has 5 irreducible representations:
 {\bf 1$_1$}, {\bf 1$_2$}, {\bf 2},  {\bf 3$_1$}, {\bf 3$_2$}.
The two singlets differ because the  {\bf 1$_1$} is invariant under the full
 group $S_4$ while  {\bf 1$_2$} changes sign  for odd permutations. Also 
 {\bf 3$_1$} and  {\bf 3$_2$} are different. {\bf 3$_1$} contains a 
$S_3\subset S_4$ singlet, while  {\bf 3$_2$} contains an $A_3\subset S_4$ 
invariant singlet.
  A$_3$ is the group of even permutations of three objects. 
The full set of  fermion families is contained in the smallest (non trivial) 
representation of E$_6^4 >{\hspace{-6pt}}\triangleleft$ S$_4$.
This is the {\bf 108}.
The product of two {\bf 108}  give the following irreducible representations
\begin{equation}
 {\bf 108}\times {\bf 108}={\bf 108}+{\bf 1404_s}+{\bf 1404_a}+
{\bf 4374_s}+{\bf 4374_a}.
\end{equation}
One can derive the following branching rules in the embedding 
 E$_6^4 >{\hspace{-6pt}}\triangleleft$ S$_4$ $\supset E_6^4$ 
\begin{small}
\begin{eqnarray*}
{\bf 108}&=&
({\bf 27},{\bf 1},{\bf 1},{\bf 1})+({\bf 1},{\bf 27},{\bf 1},{\bf 1})
+({\bf 1},{\bf 1},{\bf 27},{\bf 1})+({\bf 1},{\bf 1},{\bf 1},{\bf 27})\\
{\bf 1404_s}&=&({\bf 351^\prime},{\bf 1},{\bf 1},{\bf 1})+({\bf 1},
{\bf 351^\prime},{\bf 1},
{\bf 1})+({\bf 1},{\bf 1},{\bf 351^\prime},{\bf 1})+({\bf 1},{\bf 1},{\bf 1},
{\bf 351^\prime})\\
{\bf 4374}&=& ({\bf 27},{\bf 27},{\bf 1},{\bf 1})
+({\bf 27},{\bf 1},{\bf 27},{\bf 1})
+({\bf 27},{\bf 1},{\bf 1},{\bf 27})+({\bf 1},{\bf 27},{\bf 27},{\bf 1})\\
&+&({\bf 1},{\bf 27},{\bf 1},{\bf 27})+({\bf 1},{\bf 1},{\bf 27},{\bf 27})
\\
{\bf 6912}&=&({\bf 1728},{\bf 1},{\bf 1},{\bf 1})+({\bf 1},{\bf 1728},
{\bf 1},{\bf 1})+({\bf 1},{\bf 1},{\bf 1728},{\bf 1})+({\bf 1},{\bf 1},{\bf 1},
{\bf 1728}) \\ 
{\bf 9720 }&=&({\bf 2430},{\bf 1},{\bf 1},
{\bf 1})+({\bf 1},{\bf 2430},
{\bf 1},{\bf 1})+({\bf 1},{\bf 1},{\bf 2430},{\bf 1})+({\bf 1},{\bf 1},{\bf 1},
{\bf 2430}). 
\end{eqnarray*}
\end{small}
The standard model Higgs doublet $H$ is contained  in the {\bf 1404$_s$}.
Note that the {\bf 351$^\prime$} of E$_6$ 
contained in this representation  is the symmetric product of two 
{\bf 27} of the E$_6$ group. The {\bf 351$^\prime$}
 contains a doublet with the same quantum 
numbers of the standard model Higgs. In fact the   {\bf 351$^\prime$} of $E_6$ 
contains the ({\bf 1}, {\bf 2}, { +1/2},{ -3},{ -5}) under
 $SU_c(3)\times SU_L(2)\times U_Y(1)\times U_r(1) \times U_t(1)$ 
(see the discussion  below eq.(\ref{down}))
 that is all quantum numbers needed  
for the Higgs chosen in our model.
The scalar field $\omega_{ij}$ introduced in eq.(\ref{15}) belongs to 
the {\bf 4374$_s$} of  
E$_6^4 >{\hspace{-6pt}}\triangleleft$ S$_4$. 

The six independent components of the antisymmetric tensor
 $\omega_{ij}$ discussed in the paper, transform as the 
 {\bf 6$_a$}, that is an  irreducible representation 
 of  $(U(1)_{\mathrm{down}})^4$ $>{\hspace{-6pt}}\triangleleft$ S$_4$. They 
 are  embedded in the {\bf 4374$_s$}.
The field  $\phi^i$ that  appears in the Yukawa interactions of the down quark 
 in eq.(\ref{eq7}) belongs to   the {\bf 6912}.
The fields $\xi^i$, $\chi^i$ giving mass to the neutrino sector belong to 
the   {\bf 1404$_s$}. Finally the  {\bf 9720 } contains a 
scalar field $\phi^u_i$ that is a standard model  singlet, with the right 
quantum numbers needed  to  introduce  the Yukawa interactions for the up 
quarks.
\subsection*{Appendix B. The neutrino  mass matrix} 
The neutrino 12$\times$12  mass matrix $M_\nu$ 
coming from (\ref{eq.18}) can be 
written
\begin{small}
$$
\pmatrix{ 0 & 0 & 0 & 0 & g\,
   H & 0 & 0 & 0 & 0 & 0 & 0 & 0 \cr 0 & 0 & 0 & 0 & 0 & g\,
   H & 0 & 0 & 0 & 0 & 0 & 0 \cr 0 & 0 & 0 & 0 & 0 & 0 & g\,
   H & 0 & 0 & 0 & 0 & 0 \cr 0 & 0 & 0 & 0 & 0 & 0 & 0 & g\,
   {H_4} & 0 & 0 & 0 & 0 \cr g\,H & 0 & 0 & 0 & {g_2}\,
   {v_1} & 0 & 0 & 0 & 0 & 0 & 0 & {g_1}\,v \cr 0 & g\,
   H & 0 & 0 & 0 & {g_2}\,{v_2} & 0 & 0 & 0 & 0 & 0 & 
   {g_1}\,v \cr 0 & 0 & g\,H & 0 & 0 & 0 & {g_2}\,
   {v_2} & 0 & 0 & 0 & 0 & {g_1}\,v \cr 0 & 0 & 0 & g\,
   {H_4} & 0 & 0 & 0 & {g_2}\,{v_4} & - {g_1}\,v   & -{g_1}\,v   & -
 {g_1}\,v   & 0 \cr 0 & 0 & 0 & 0 & 0 & 0 & 0 & -
 {g_1}\,v   & {g_3}\,
   {v_1^\prime} & 0 & 0 & 0 \cr 0 & 0 & 0 & 0 & 0 & 0 & 0 & - 
     {g_1}\,v  & 0 & {g_3}\,
   {v_2^\prime} & 0 & 0 \cr 0 & 0 & 0 & 0 & 0 & 0 & 0 & - 
     {g_1}\,v   & 0 & 0 & {g_3}\,
   {v_2^\prime} & 0 \cr 0 & 0 & 0 & 0 & {g_1}\,v & 
   {g_1}\,v & {g_1}\,v & 0 & 0 & 0 & 0 & {g_3}\,
   {v_4^\prime} \cr  \nonumber} \nonumber
$$
\end{small}
in the following  flavor basis 
\begin{equation}
v=\left(\nu_L^1,\nu_L^2,\nu_L^3,\nu_L^4,X_L^1,X_L^2,X_L^3,X_L^4,
\nu_R^1,\nu_R^2,\nu_R^3,\nu_R^4\right).
\end{equation}

To find the eigenvectors and the eigenvalues of the matrix $M_\nu$ of the 
three lightest neutrinos we proceed as follows.
First we show that the following vector $v_1$  is a light eigenstate of the 
matrix $M_\nu$  
\begin{equation}
v_1
=\left(\sqrt{\frac{2}{3}}, \frac{-1}{\sqrt{6}}, \frac{-1}{\sqrt{6}}, 0, 
z \sqrt{\frac{2}{3}} + x , \frac{-z}{\sqrt{6}} + x , \frac{-z}{\sqrt{6}}
 + x , 0 , 0, 0, 0,
 y\right).
\end{equation}
In fact, for very small values of $m_1,x,y,z$ and taking into account that 
$v\gg v_1, v_2 ,v_4^\prime $ we have that 
\begin{equation}\label{eq30}
 M_\nu v_1= m_1 v_1^0
\end{equation}
where $v_1^0=v_1$ when $x=y=z=0$. Eq.(\ref{eq30}) holds only at the first 
order of 
the perturbative expansion in terms of the unknown variables $m_1,x,y,z$.
It is not difficult to solve the system of equations given in eq.(\ref{eq30}).
Namely, Eq.(\ref{eq30})  gives  a linear system of three equations with 
respect the  
three variables $x,y,z$.    
If we put the solution of this system for the three variables $x,y,z$ into
the eq.(\ref{eq30}) we can also derive the eigenvalue $m_1$ that corresponds 
to the
 mass of the neutrino with the following components
\begin{equation} 
v_1^0=\left(\sqrt{2/3}, -1/\sqrt{6}, -1/\sqrt{6}, 0,0 ,0,0,0,0,0, 0, 0
\right)
\end{equation}
and 
\begin{equation}
m_1= g H \left(\sqrt{6}\, x-z\right)=\frac{3\,g^2\,H^2}
  {{g_2}\,\left( 2\,{v_1} + {v_2}
      \right) }
\end{equation}
where in last step we have used that $v\gg v_1,v_2,v_4^\prime$.
In a similar way, one can get the remaining two light eigenstates with the 
corresponding eigenvalues
\begin{eqnarray}
&& v_2^0
= \left(1/\sqrt{3}, 1/\sqrt{3}, 1/\sqrt{3}, 0,0 ,0,0 , 0 , 0, 0, 0, 0\right)
 \\
&& v_3^0
= \left(0, 1/\sqrt{2}, -1/\sqrt{2}, 0,0 ,0,0 , 0 , 0, 0, 0, 0\right)
 \\
&&m_2=\frac{g_3 g^2 H^2 v_4^\prime}{3 g_1^2 v^2}\\ 
&&m_3=\frac{g^2 H^2}{ g_2 v_2}.
\end{eqnarray}

\end{document}